\documentclass[letterpaper, 12pt]{article}

\title{Two--Sample Inference in Highly Dispersed Negative Binomial Models}
\author{David Shilane and Derek Bean}

\usepackage{graphicx} 
\usepackage{amsmath}  
\usepackage{amssymb}
\usepackage[sort&compress]{natbib}
\usepackage{geometry}
\begin{document}
\maketitle
\pagestyle{empty}

\section{Introduction}
\label{intro}

\noindent Given independent random samples of data, the difference in sample means is a common measure of disparity between two populations.  When the sample sizes are large and the samples' respective distributions are reasonably well--behaved, a Normal distribution approximation to the mean difference is commonly employed.  This approximation is justified by the Central Limit Theorem.  However, in practice it is often difficult to determine the required sample sizes needed to ensure a reliable inference.  Data arising from a highly dispersed Negative Binomial model may be extremely skewed.  In many cases, a one--sample Normal approximation for a Negative Binomial mean does not provide reliable estimates, even at sample sizes typically considered sufficiently large (e.g. $n=50$ or 100).\\

\noindent \citet{Shilane-negative-binomial-2010} investigated alternative methods for one--sample inference in highly dispersed Negative Binomial models.  These methods include a bootstrap approach, tail probability bounds such as Bernstein's Inequality, and parametric methods based upon the Normal, Gamma, and Chi Square distributions.  We seek to extend this analysis to the two--sample case.  When each sample mean would best be approximated with either a Normal, Gamma, or Chi Square distribution, we will demonstrate that a Normal approximation is appropriate for two--sample inferences.  We will also adapt Bernstein's Inequality to generate inferences in two--sample cases for which the Normal approximation and Bootstrap methods are unreliable.

\section{One--Sample Inference in Negative Binomial Models}
\label{onesample}

\subsection{The Negative Binomial Distribution}
\label{negbindist}

\noindent A Negative Binomial variable $X$ typically models the random number of failures $k\in\mathbb{Z}^+$ observed before the $r$th success (with $r\in\mathbb{Z}^+$) over a series of trials.  Each trial is the result of an independent, identically distributed (i.i.d.) Bernoulli random variable that results in success with probability $p$ and failure otherwise.  The Negative Binomial distribution may be characterized in terms of the parameters $r$ and $p$.  An alternative parameterization sets a mean parameter $\mu \equiv r \left(\frac{1}{p} - 1\right)$ and a dispersion parameter $\theta \equiv r$.  We will adopt this parameterization for the remainder of this study.  The probability mass function \citep{Hilbe} of the Negative Binomial NB($\mu,\theta$) random variable $X$ is then given by

\begin{equation}\label{PMFP4}
P(X=k) = \frac{\mu^k}{k!} \frac{\Gamma(\theta+k)}{\Gamma(\theta)[\mu+\theta]^k}\frac{1}{\left(1+\frac{\mu}{\theta}\right)^\theta}, k\in\mathbb{Z}^+.
\end{equation}

\noindent The Negative Binomial distribution serves as a general model for i.i.d. counting data $X_1, \dots, X_n$ with $n\in\mathbb{Z}^+$.  The Poisson distribution, which is often used to model counts, corresponds to the special case of $\theta\rightarrow\infty$.  Under a Poisson model, the mean and variance are equal.  For any finite value of $\theta$, the variance of a Negative Binomial is greater than the mean.  This dispersion grows as $\theta$ decreases.  When $\theta$ is small, the distribution becomes highly skewed.  \citet{Shilane-negative-binomial-2010} demonstrate that the sample mean of i.i.d. Negative Binomial random variables exhibits a slow convergence to the Normal distribution.  As such, a Central Limit Theorem approximation may perform poorly at moderate samples sizes (e.g. $n=50$ or 100).

\subsection{Inference}
\label{inference}

\begin{table}[ht] 
\begin{center}
\begin{tabular}{|c|c|}
\hline
\textbf{Scenario} & \textbf{Preferred Method}\\
\hline
Small $n$, small $\theta$ & Bounded Bernstein\\
\hline
Large $n$, small $\theta$ & Gamma\\
\hline
Small $n$, large $\theta$ & Normal, Bootstrap, or Bounded Bernstein\\
\hline
Large $n$, large $\theta$ & Normal or Bootstrap\\
\hline
$\mu\approx 2n\theta$ & Chi Square\\
\hline
\end{tabular}
\caption{General guidelines for selecting among the proposed methods for one--sample inference in Negative Binomial models.}
\label{guidelines}
\end{center}
\label{default}
\end{table}

\noindent Because the Normal approximation cannot ensure reliable estimates of the mean $\mu$, \citet{Shilane-negative-binomial-2010} proposed a variety of methods for one--sample inference in highly skewed Negative Binomial models.  These include approximations based upon the Gamma and Chi Square distributions, the Bootstrap Bias--Corrected and Accelerated (BCA) method \citep{Bootstrap}, and tail probability bounds such as Bernstein's Inequality.  The proposed methods are largely complementary.  Table \ref{guidelines} provides guidelines for selecting an appropriate method according to the scenario.  The exact boundary at which one method overtakes another depends upon the sample size $n$ along with the parameters $\mu$ and $\theta$.

\section{Methods}
\label{methods}

\noindent We seek to provide adequate methods for two--sample inference in highly dispersed Negative Binomial Models.  The data consist of $X = \left(X_1, \dots, X_{n_x}\right)$, which are i.i.d. NB($\mu_x, \theta_x$), and $Y = \left(Y_1, \dots, Y_{n_y}\right)$, which are i.i.d. NB($\mu_y, \theta_y$).  The two samples $X$ and $Y$ are independent.  In this setting, the difference in means $\mu_x-\mu_y$ is our parameter of interest.  We will estimate this parameter with $\bar X - \bar Y$, the difference in sample means.  The methods of inference will consist of estimating the distribution of $\bar X - \bar Y$ or providing appropriate probability tail bounds using Bernstein's Inequality.\\

\noindent Inferences about $\bar X - \bar Y$ may be obtained by the Bootstrap method.  Otherwise, since inferences about $\bar X$ and $\bar Y$ may be independently approximated by tail probability bounds or any of the Gamma, Chi Square, and Normal distributions, the difference $\bar X-\bar Y$ may be categorized by 16 cases.  When a bound like Bernstein's Inequality is required for either sample individually, it will also be applied to the two--sample case.  When both sample means are approximately Normal, the standard two--sample Normal approximation may be applied.  Since the Chi Square distribution is a special case of the Gamma, the remaining cases only require ascertaining the distribution of the difference of two Gammas or that of one Gamma and one Normal.  The following subsections will adapt Bernstein's Inequality to the two--sample case and show that any difference of Gamma and Normal variables is approximately Normal.

\subsection{Bernstein's Inequality}

\noindent When at least one of the sample sizes $n_x$ or $n_y$ is sufficiently small, the distribution of the respective sample mean $\bar X$ or $\bar Y$ is not well--approximated by a Gamma or Normal distribution.  The Chi Square model applies if $\mu \approx 2n\theta$.  In all other cases, we must rely upon probability tail bounds to perform inference on $\bar X-\bar Y$.  \citet{Shilane-negative-binomial-2010} recommend a bounded variant of Bernstein's Inequality for the one--sample setting.  We will briefly review the one--sample Bernstein method and then introduce an extension for the two--sample setting.\\

\noindent Let $Z= \left(Z_1, \dots, Z_n\right)$ be independent random variables bounded in a range $(a,b)\in\mathbb{R}, a < b$.  Bernstein's Inequality \citep{Shilane-negative-binomial-2010,rosenblum&vdlaanTechRep130} states that

\begin{equation}\label{bernsteinEQ}
P\left(\frac{1}{n}\left| \sum_{i=1}^{n} (Z_i - E[Z_i]) \right| > \epsilon\right) \leq 2 \exp\left[\frac{-1}{2}\left(\frac{n\epsilon^2}{\sigma^2 + \epsilon(b-a)/3}\right)\right].
\end{equation}

\noindent When the right side of Equation (\ref{bernsteinEQ}) is set equal to $\alpha/2$, we can construct a $1-\alpha$ confidence interval for $E[Z]$.  Such an interval will have the form $\bar Z \pm \epsilon$, where $\epsilon$ is given by

\begin{equation}\label{ep}
\epsilon = \frac{\frac{-2}{3}(b-a)\log(\alpha/2)\pm \sqrt{\frac{4}{9}(b-a)^2[\log(\alpha/2)]^2 - 8n\sigma^2\log(\alpha/2)}}{2n}.
\end{equation}

\noindent The two--sample case can be adapted to the form of the one--sample version of Bernstein's Inequality.  Consider the following transformation of the data:  Let $n=n_x+n_y$, and define $Z_1, \dots, Z_n$ as

\begin{equation}\label{transformeddata}
Z_i = \begin{cases}  \frac{n}{n_x} X_i & \mbox{if } i\in\{1,\dots,n_x\}; \\
  \frac{-n}{n_y} Y_{i-n_x} & \mbox{if } i\in\{n_x+1,\dots,n\}.
  \end{cases}
\end{equation}

\noindent The data set $Z$ is constructed so that $\bar Z = \bar X - \bar Y$.  Therefore, $E[\bar Z] = \mu_x-\mu_y$ and $Var(\bar Z) = \frac{\sigma_x^2}{n_x} + \frac{\sigma_y^2}{n_y}$.  Since $Z_1,\dots, Z_n$ are independent, bounded variables, the version of Bernstein's Inequality given by Equation (\ref{bernsteinEQ}) may be applied.  The bounding range $(a,b)$ may be specified in terms of the maximum values of the two data sets.  Once the sample size $n$, variance $\sigma^2$, and bounding range $(a,b)$ are specified, Bernstein's Inequality may be applied.  These parameters are:

\begin{equation}\label{bernparams}
\begin{split}
& n = n_x + n_y; \\
& \sigma^2 = n \cdot Var(\bar Z) = \frac{n_x+n_y}{n_x}\sigma_x^2 + \frac{n_x+n_y}{n_y}\sigma_y^2;\\
& a = c_a \frac{-n}{n_y}\max\left(Y_1,\dots,Y_{n_y}\right) \texttt{with $c_a=1$ by default};\\
& b = c_b \frac{n}{n_x}\max\left(X_1,\dots,X_{n_x}\right) \texttt{with $c_b=1$ by default}.
\end{split}
\end{equation}

\noindent Applying these parameters to Equation (\ref{ep}), a $1-\alpha$ confidence interval for $\mu_x-\mu_y$ is given by $\bar X-\bar Y \pm \epsilon$.  Furthermore, a test of the null hypothesis $H_0: \mu_x-\mu_y = w$ versus the two--sided alternative $H_A: \mu_x-\mu_y \neq w$ can also be performed using Bernstein's Inequality.  In this case, the value of $\epsilon$ is given by $\bar X - \bar Y - w$.  Then the $p$-value for this test is the value of $\alpha$ solving Equation (\ref{ep}), which requires an application of the Quadratic Formula:

\begin{equation}\label{alpha}\begin{split}
\alpha = 2&\exp\left[\frac{\frac{-8}{3}n\epsilon(b-a)+\frac{4}{9}(b-a)^2 - 8n\sigma^2}
							  {\frac{8}{9}(b-a)^2}\right]\\
 *&\exp\left[\pm \frac{\sqrt{\left(\frac{8}{3}n\epsilon(b-a)  - \frac{4}{9}(b-a)^2+ 8n\sigma^2\right)^2 - \frac{64}{9}n^2\epsilon^2(b-a)^2}}{\frac{8}{9}(b-a)^2}\right].
\end{split}\end{equation}

\noindent One caveat to the proposed use of Bernstein's Inequality is that Negative Binomial variables are in fact unbounded above.  Any selected bounding range $(a,b)$ will be at best a heuristic assumption.  \citet{Shilane-negative-binomial-2010} considered both bounded \citep{rosenblum&vdlaanTechRep130} and unbounded \citep{BirgeMassart} variants of Bernstein's Inequality.  The Bounded Bernstein method for one--sample inference proved to be a useful tool at small sample sizes in simulation studies.  However, the Unbounded Bernstein method was not able to generate inferences of a reasonable quality because its tail probability bound was not sufficiently sharp. There are limited guidelines for selecting $(a,b)$.  At minimum, the respective samples' maximum values could be selected; that is, the constants $c_a$ and $c_b$ should be at least one.\\  

\noindent A variety of other tail probability bounds may be employed in place of Bernstein's Inequality.  These include other varieties of Bernstein's Inequality \citep{Bernstein}, Bennett's Inequality \citep{Bennett1962, Bennett1963}, Hoefding's Method \citep{Hoeffding}, McDiarmid's Inequality \citep{mcdiarmid,kutin}, and the Berry-Esseen Inequality \citep{Berry, Esseen1942, Esseen1956, vanBeek}.

\subsection{Parametric Approaches}

\noindent When both samples' respective means can be modeled with either a Chi Square, Gamma, or Normal distribution, the difference is sample means will be approximately Normal.  We can establish this by considering the Laplace transform of each possible pair of distributions.  As an example, suppose $\bar X$ is approximately Gamma$\left(n_x\theta_x, \frac{n_x\theta_x}{\mu_x}\right)$ and  $\bar Y$ is approximately Gamma$\left(n_y\theta_y, \frac{n_y\theta_y}{\mu_y}\right)$.  Then the Laplace transform of $\bar X - \bar Y$ is:

\begin{equation}\label{gamma-gamma}
L_{\bar X - \bar Y}(\lambda) = L_{\bar X}(\lambda) L_{\bar Y}(-\lambda) =  \left(1-\frac{\mu_x\lambda}{n_x\theta_x}\right)^{-n_x\theta_x} \left(1+\frac{\mu_y\lambda}{n_y\theta_y}\right)^{-n_y\theta_y}.
\end{equation}

\noindent The natural logarithm of this transform is then:

\begin{equation}\label{lg-gamma-gamma}
 \log \left(L_{\bar X - \bar Y}(\lambda)\right)= -n_x\theta_x \log\left(1-\frac{\mu_x\lambda}{n_x\theta_x}\right)  -n_y\theta_y\log\left(1+\frac{\mu_y\lambda}{n_y\theta_y}\right).
\end{equation}

\noindent Using the first and second--order Taylor series approximation $\log(1+v) \approx v - \frac{v^2}{2}$, Equation (\ref{lg-gamma-gamma}) is approximately:

\begin{eqnarray}\label{log-gamma-gamma}
 \log \left(L_{\bar X - \bar Y}(\lambda)\right) &\approx&  -n_x\theta_x \left(\frac{-\mu_x\lambda}{n_x\theta_x} -\frac{\mu^2_x\lambda^2}{2n^2_x\theta^2_x}\right) -n_y\theta_y \left(\frac{\mu_y\lambda}{n_y\theta_y} -\frac{\mu^2_y\lambda^2}{2n^2_y\theta^2_y}\right) \nonumber\\ 
 &=&  (\mu_x-\mu_y)\lambda + \left(\frac{\mu_x^2}{n_x\theta_x} + \frac{\mu_y^2}{n_y\theta_y}\right) \frac{\lambda^2}{2}.
 \end{eqnarray}

\noindent Exponentiating both sides of Equation (\ref{log-gamma-gamma}) shows that the Laplace transform of $\bar X-\bar Y$ has an approximately Normal distribution with mean $\mu_x-\mu_y$ and variance $\frac{\mu_x^2}{n_x\theta_x} + \frac{\mu_y^2}{n_y\theta_y}$.  That is,  $\bar X-\bar Y \approx N\left(\mu_x-\mu_y, \frac{\mu_x^2}{n_x\theta_x} + \frac{\mu_y^2}{n_y\theta_y}\right)$.\\

\begin{table}[htdp]
\begin{center}
\begin{tabular}{|c|c|c|c|}
\hline
 & \begin{tiny}\textbf{Normal} \end{tiny} & \begin{tiny}\textbf{Gamma}  \end{tiny} & \begin{tiny} \textbf{Chi Square} \end{tiny}\\
 \hline
 \begin{tiny}\textbf{Normal} \end{tiny}& \begin{tiny}$N\left(\mu_x-\mu_y, \frac{\sigma_x^2}{n_x} + \frac{\sigma_y^2}{n_y}\right)$ \end{tiny}&\begin{tiny} $N\left(\mu_x-\mu_y, \frac{\mu_x(\mu_x+\theta_x)}{n_x\theta_x} + \frac{\mu_y^2}{n_y\theta_y}\right)$\end{tiny} & \begin{tiny} $N\left(\mu_x-\mu_y, \frac{\mu_x(\mu_x+\theta_x)}{n_x\theta_x} + 2\mu_y\right)$\end{tiny}\\
 \hline
 
 \begin{tiny}\textbf{Gamma}  \end{tiny}& \begin{tiny} $N\left(\mu_x-\mu_y, \frac{\mu_x^2}{n_x\theta_x} + \frac{\mu_y(\mu_y+\theta_y)}{n_y\theta_y}\right)$ \end{tiny} & \begin{tiny} $N\left(\mu_x-\mu_y, \frac{\mu_x^2}{n_x\theta_x} + \frac{\mu_y^2}{n_y\theta_y}\right)$ \end{tiny} & \begin{tiny} $N\left(\mu_x-\mu_y, \frac{\mu_x^2}{n_x\theta_x} + 2\mu_y\right)$\end{tiny}\\
 \hline

\begin{tiny} \textbf{Chi Square} \end{tiny} & \begin{tiny} $N\left(\mu_x-\mu_y, 2\mu_x + \frac{\mu_y(\mu_y+\theta_y)}{n_y\theta_y}\right)$\end{tiny} & \begin{tiny} $N\left(\mu_x-\mu_y, 2\mu_x + \frac{\mu_y^2}{n_y\theta_y}\right)$ \end{tiny} & \begin{tiny} $N\left(\mu_x-\mu_y, 2(\mu_x+\mu_y)\right)$\end{tiny}\\
 \hline
 \end{tabular}
\end{center}
\caption{The distribution of $\bar X - \bar Y$.  The rows represent $\bar X$ and the columns $\bar Y$.}
\label{methods_summary}
\end{table}

\noindent Similar arguments may be applied when one of $\bar X$ or $\bar Y$ is approximately Normal and the other is Gamma or Chi Square.  The difference $\bar X - \bar Y$ will be approximately Normal for all 9 parametric combinations.  The parameters of these Normal distributions are given in Table \ref{methods_summary}.  In all other circumstances, inference may be obtained using the Bootstrap method or an appropriate tail probability bound such as Bernstein's Inequality.\\

\noindent In all cases, the mean difference $\mu_x-\mu_y$ is estimated by the statistic $\bar X-\bar Y$.  The variance of the sample mean difference depends upon the mean and dispersion parameters of the two samples.  Two estimation methods may be considered.  One approach would consist of first estimating the dispersion parameters $\theta_x$ and $\theta_y$ and then plugging these estimates into the appropriate scenario in Table \ref{methods_summary}.  The second approach is to directly estimate the sample variances with the statistics $s_x^2$ and $s_y^2$.  This approach directly estimates the variance parameter without relying upon estimates of the nuisance parameters $\theta_x$ and $\theta_y$.  Therefore, the estimated variance of $\bar X-\bar Y$ is the familiar $s_x^2/n_x + s_y^2/n_y$.\\

\noindent We recommend the latter approach of directly estimating $s_x^2$ and $s_y^2$, especially in light of the difficulty of estimating small values of $\theta_x$ and $\theta_y$.  These dispersion parameters can be estimated through the method of moments \citep{Pietersetal,Shilane-negative-binomial-2010} or numeric maximum likelihood estimation (MLE) procedures \citep{Piegorsch,ClarkPerry}.  \citep{Pietersetal} provides a comparison of these procedures.  However, the MLE does not necessarily exist \citep{Aragonetal,Ferreri}.  In practice, MLE estimates are either highly variable or generate computational errors in software implementations when the dispersion is very small.  Meanwhile, the method of moments estimator results in negative estimates of the strictly positive dispersion when the data's sample variance is less than the sample mean.  Even if these difficulties were resolved, direct estimation is typically more efficient than plug--in estimators.  With these considerations in mind, we will rely upon direct estimates of the sample variances and avoid unnecessary estimation of the dispersion parameters.
 
 \subsection{A Mixture Method}\label{mixture}
\noindent In general, we expect the Bernstein method to produce more conservative and considerably wider confidence intervals than the Normal approximation.  As such, these techniques may be used in a complementary fashion.  When the sample sizes are small and the dispersion is high, Bernstein confidence intervals will be more reliable.  At larger sample sizes and more moderate dispersions, the Normal approximation should be sufficient.  We also propose a Mixture method that averages the lower ($L$) and upper ($U$) end--points of the intervals.  Such a method may produce improvements in boundary settings in which the Normal approximation is gaining in reliability but still insufficient for inference.  Other weighted combinations may be considered of the form

\begin{equation}\label{mixeq}
\begin{split}
\left(L_{\texttt{Mixture}}, U_{\texttt{Mixture}}\right)  & = w\left(L_{\texttt{Normal}}, U_{\texttt{Normal}}\right)\\
&  + (1-w)\left(L_{\texttt{Bernstein}}, U_{\texttt{Bernstein}}\right); w\in[0,1].\\
\end{split}
\end{equation}

\noindent We will set $w=0.5$ as a default, which corresponds to the case of averaging the Normal and Bernstein intervals.

\section{Simulation Studies}\label{sim}

\begin{table}[ht]
\begin{center}
\begin{tabular}{|c|c|}
\hline
\textbf{Parameter} & \textbf{Values}\\
\hline
$\mu_x$ & $\{5, 10\}$\\
\hline
$\mu_y$ & $\{5, 10\}$\\
\hline
$\theta_x$ & $\{0.01, 0.025, 0.05, 0.075, 0.1\}$\\
\hline
$\theta_y$ & $\{0.01, 0.025, 0.05, 0.075, 0.1\}$\\
\hline
$n_x$ & $\{10, 20, 30, \dots, 180, 190, 200, 250, 500, 1000\}$\\
\hline
$n_y$ & $\{10, 20, 30, \dots, 180, 190, 200, 250, 500, 1000\}$\\
\hline
Trials & 10000\\
\hline
\end{tabular}
\caption{Parameter values for the simulation experiments of Section \ref{sim}.  Each choice of sample sizes $n_x$ and $n_y$, means $\mu_x$ and $\mu_y$, and dispersions $\theta_x$ and $\theta_y$ comprised an independent simulation experiment.  A total of 10000 confidence intervals were randomly generated for each experiment.  Coverage probabilities were estimated by the empirical proportion of confidence intervals containing the true mean difference $\mu_x-\mu_y$.}
\label{simparams}
\end{center}
\end{table}

\noindent We assessed the quality of the proposed Normal, Bernstein, and Mixture confidence intervals in a simulation study.  We selected a wide array of two--sample inference problems in highly dispersed Negative Binomial models.  The parameter values for this simulation, which are summarized in Table \ref{simparams}, include a variety of sample sizes from small to large at dispersions ranging from large to extremely high over several combinations of means.  Each choice of sample sizes $n_x$ and $n_y$, means $\mu_x$ and $\mu_y$, and dispersions $\theta_x$ and $\theta_y$ comprised an independent simulation experiment.  Each experiment randomly generated a total of 10000 pairs of data sets including $n_x$ i.i.d. NB($\mu_x,\theta_x$) and $n_y$ i.i.d. NB($\mu_y,\theta_y$) random variables.  With $\alpha=0.05$, $95\%$ confidence intervals for the mean difference $\mu_x-\mu_y$ were constructed on each of the 10000 pairs of data sets according to the Normal, Bernstein, and Mixture methods of the previous section.  The method's coverage probability in an experiment was estimated by the empirical proportion of the 10000 confidence intervals that contained the true mean difference $\mu_x-\mu_y$.  The standard error for this estimate is given by $\sqrt{\frac{p_c(1-p_c)}{10000}}$, where $p_c$ is the true coverage probability.  When $p_c=0.95$, the 10000 repetitions ensure that the estimated coverage has a margin of error of approximately $0.004=0.4\%$.  Under the most extreme case of $p_c=0.5$, this margin of error would be approximately $0.01=1\%$.\\

\noindent This coverage probability estimation procedure was repeated across the 52900 simulation experiments defined by all unique combinations of parameters values among those listed in Table \ref{simparams}.  The Bootstrap method was not employed in this simulation because of its heavy computational burden.  Each experiment entailed the generation of $10000(n_x+n_y)$ random variables.  Over the 52900 experiments, this amounted to a total of approximately $7.84\cdot 10^{15}$ random numbers.  All told, the simulation required approximately two days of continuous computation to ascertain the quality of the Bernstein, Normal, and Mixture methods.  If the Bootstrap method were included, this would roughly increase the total random numbers to be generated in any experiment by a factor of $B(n_x+n_y)$.  If $B$ were set to 10000 or more to ensure reliable Bootstrap inferences, this simulation would be considered intractable.\\

\begin{figure}
  \includegraphics[scale=0.78]{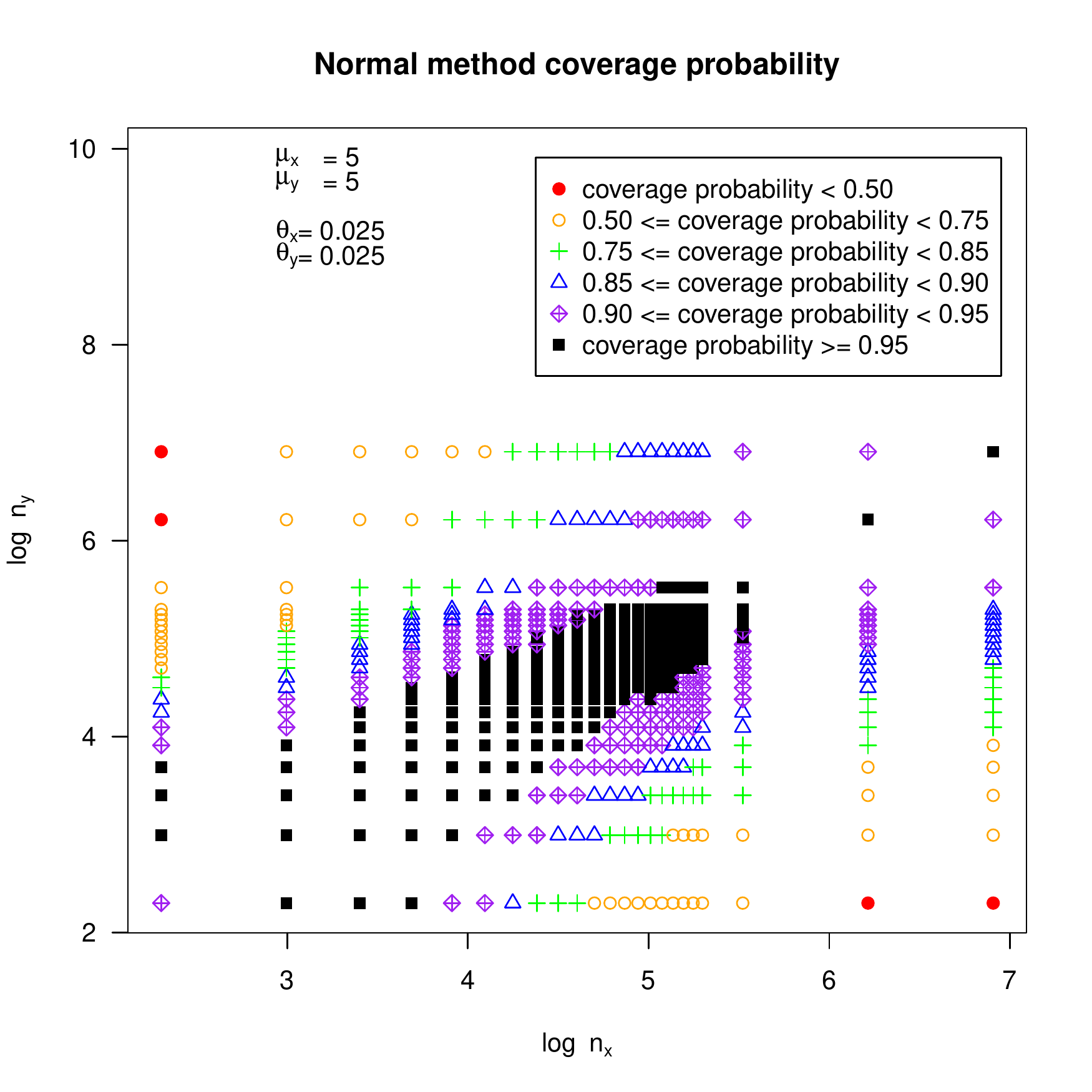}
  \caption{Simulation results for the Normal approximation with $\mu_x=\mu_y=5$ and $\theta_x=\theta_y=0.025$ across all considered sample size combinations.  These results may be directly compared to those of the Bernstein method in Figure \ref{bernsteinsim} or the Mixture Method in Figure \ref{mixturesim}.}  \label{normsim}
\end{figure}

\begin{figure}
  \includegraphics[scale=0.78]{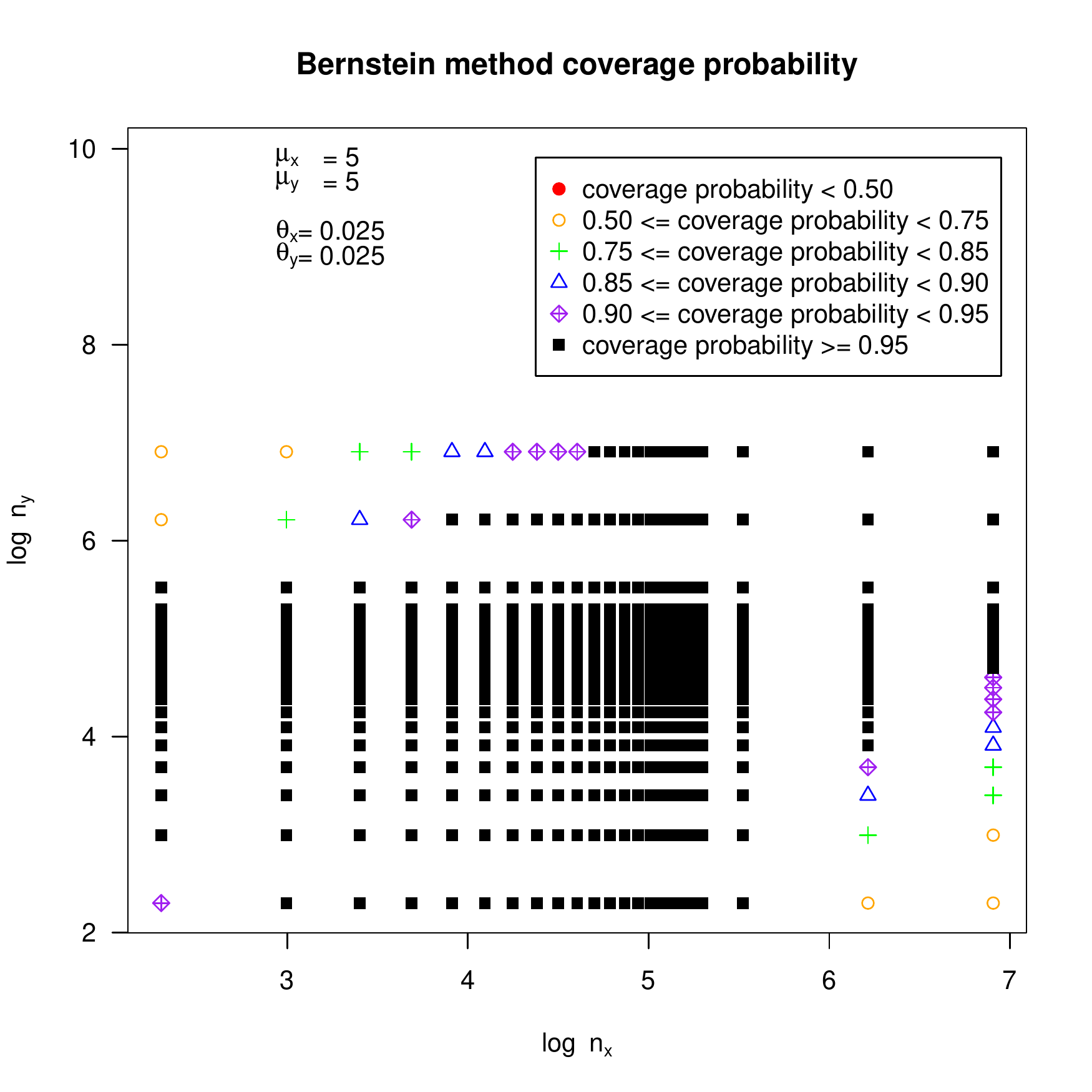}
  \caption{Simulation results for the Bernstein method with $\mu_x=\mu_y=5$ and $\theta_x=\theta_y=0.025$ across all considered sample size combinations.  These results may be directly compared to those of the Normal approximation in Figure \ref{normsim} or the Mixture method in Figure \ref{mixturesim}.}  \label{bernsteinsim}
\end{figure}

\begin{figure}
  \includegraphics[scale=0.78]{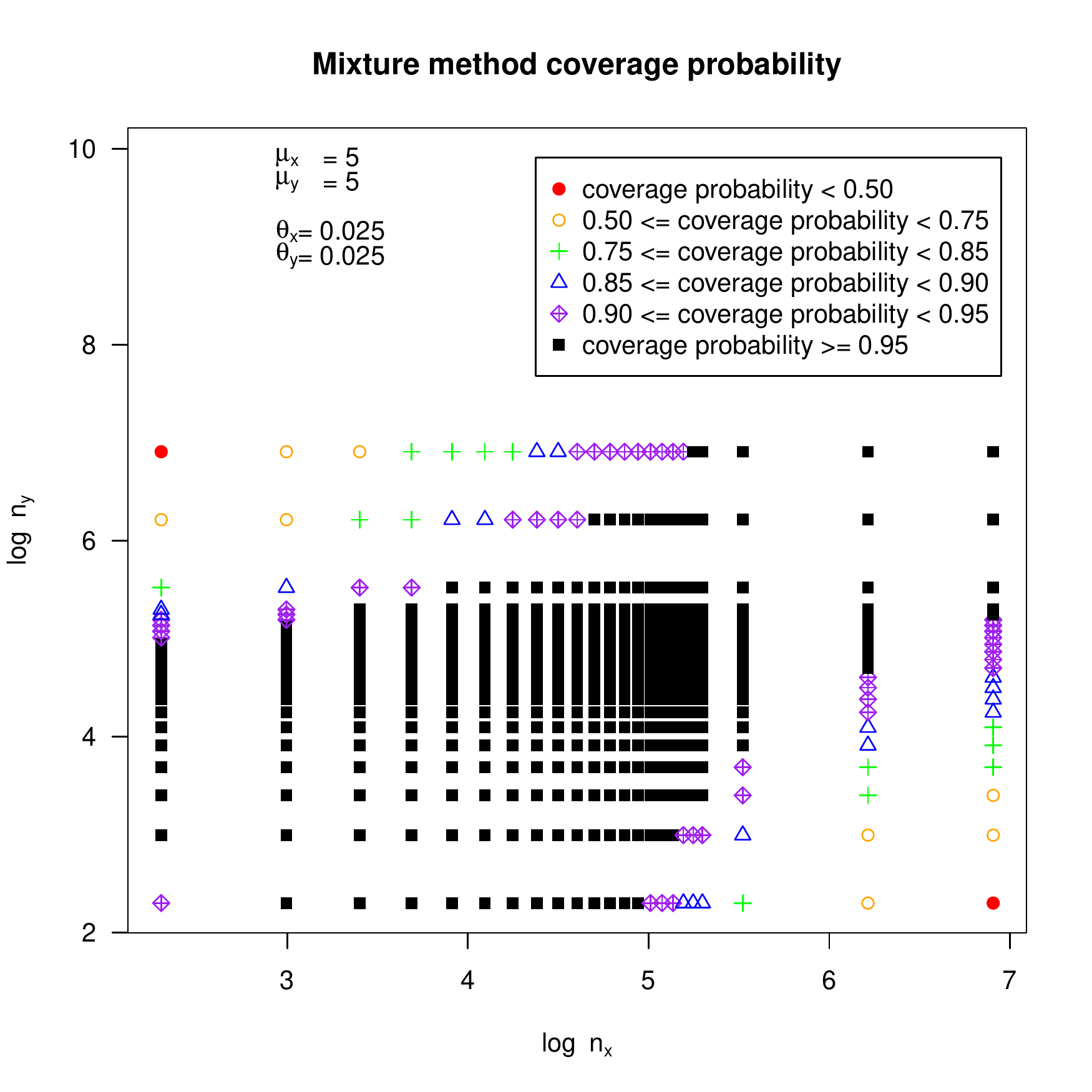}
  \caption{Simulation results for the Mixture method with $\mu_x=\mu_y=5$ and $\theta_x=\theta_y=0.025$ across all considered sample size combinations.  These results may be directly compared to those of the Normal approximation in Figure \ref{normsim} or the Bernstein method in Figure \ref{bernsteinsim}.}  \label{mixturesim}
\end{figure}

\begin{figure}
  \includegraphics[scale=0.78]{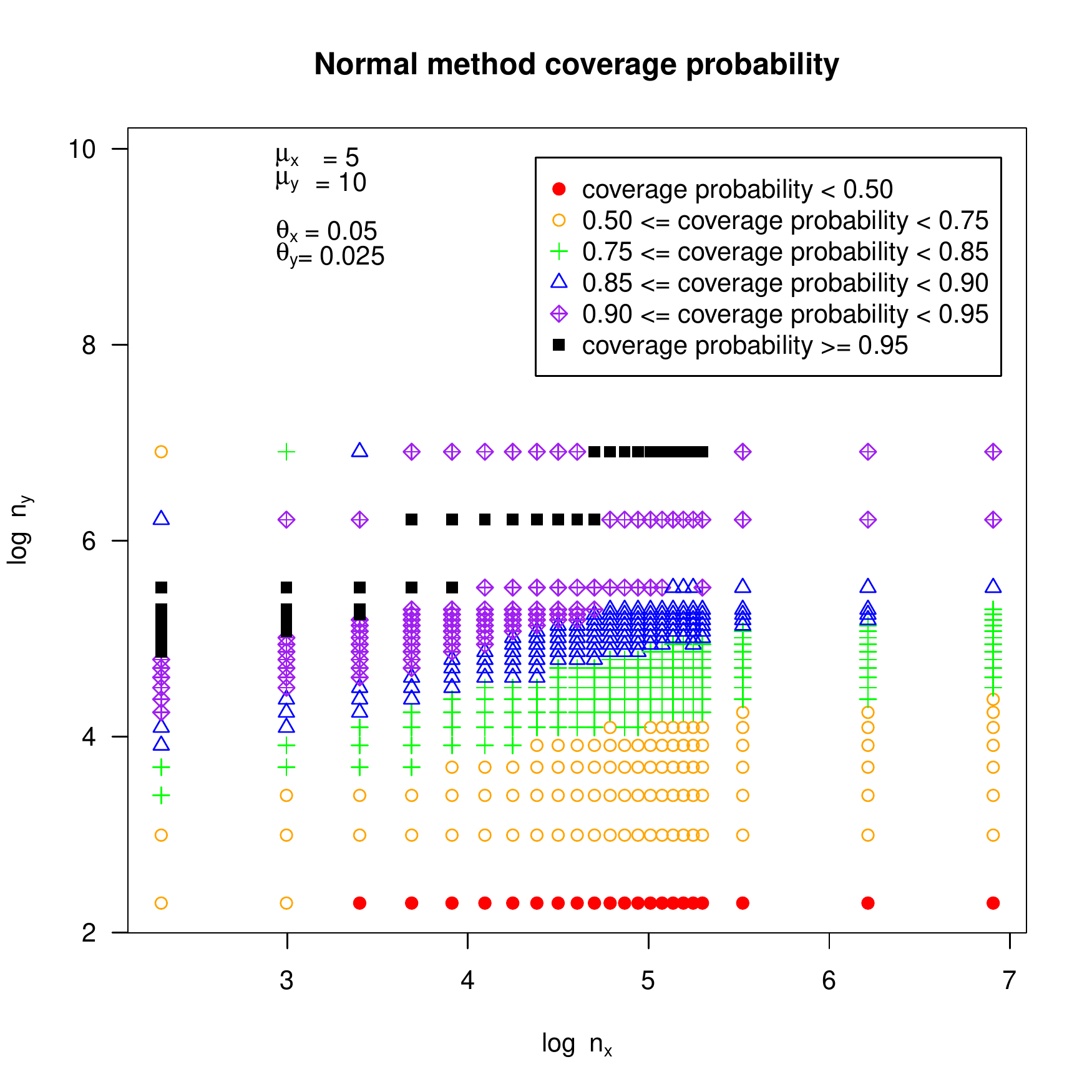}
  \caption{Simulation results for the Normal approximation with $\mu_x=5, \mu_y=10, \theta_x=0.05$, and $\theta_y=0.025$ across all considered sample size combinations.  These results may be directly compared to those of the Bernstein method in Figure \ref{bernsteinsim2} or the Mixture Method in Figure \ref{mixturesim2}.}  \label{normsim2}
\end{figure}

\begin{figure}
  \includegraphics[scale=0.78]{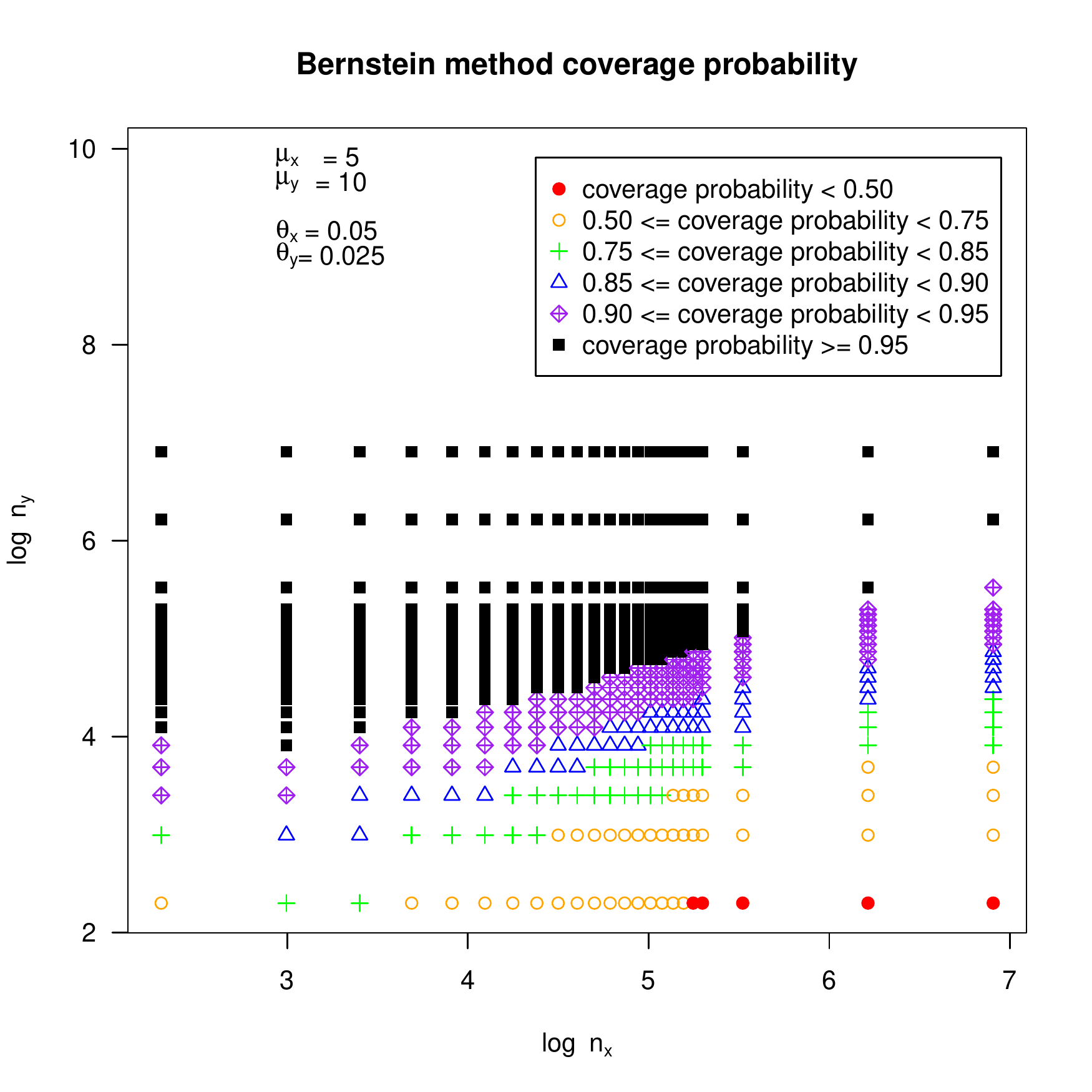}
  \caption{Simulation results for the Bernstein method with $\mu_x=5, \mu_y=10, \theta_x=0.05$, and $\theta_y=0.025$ across all considered sample size combinations.  These results may be directly compared to those of the Normal approximation in Figure \ref{normsim2} or the Mixture method in Figure \ref{mixturesim2}.}  \label{bernsteinsim2}
\end{figure}

\begin{figure}
  \includegraphics[scale=0.78]{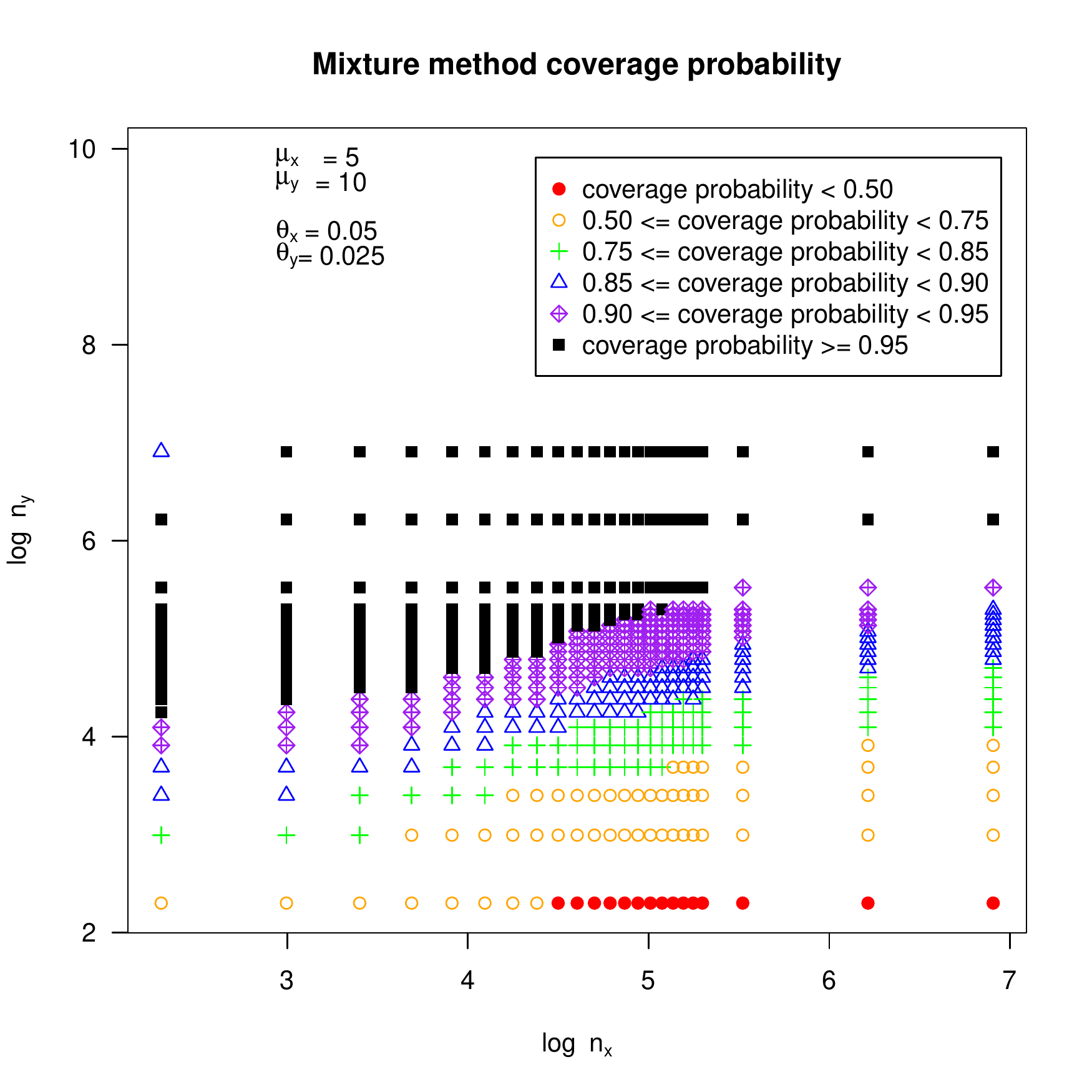}
  \caption{Simulation results for the Mixture method with $\mu_x=5, \mu_y=10, \theta_x=0.05$, and $\theta_y=0.025$ across all considered sample size combinations.  These results may be directly compared to those of the Normal approximation in Figure \ref{normsim2} or the Bernstein method in Figure \ref{bernsteinsim2}.}  \label{mixturesim2}
\end{figure}

\begin{table}[ht]
\begin{center}
\begin{tabular}{rrrrrrr}
  \hline
 & Min & 1st Qu. & Median & Mean & 3rd Qu. & Max \\ 
  \hline
Bernstein & 3.58 & 20.86 & 28.13 & 29.24 & 35.88 & 69.24 \\ 
  Mixture & 2.87 & 16.41 & 22.19 & 22.90 & 28.03 & 53.30 \\ 
  Normal & 2.16 & 11.92 & 16.13 & 16.55 & 20.18 & 37.60 \\ 
  Bernstein - Normal & 1.42 & 8.85 & 11.96 & 12.69 & 15.71 & 33.51 \\ 
   \hline
\end{tabular}
\caption{Summary of median length of each method's confidence interval across all simulation experiments.  The Bernstein - Normal row provides summary information for the length difference of the two intervals.}
\label{lengthtab}
\end{center}
\end{table}

\noindent As an example, Figures \ref{normsim}, \ref{bernsteinsim}, and \ref{mixturesim} provide coverage probabilities for all combinations of sample sizes in the case of $\mu_x=\mu_y=5$ and $\theta_x=\theta_y=0.025$.  The Normal results in Figure \ref{normsim} exhibit accurate coverage over a large portion of the sample sizes considered.  The Normal approximation performs best in a region surrounding the main diagonal, and its coverage only drops off as the disparity between the two sample sizes grows.  In cases of differing dispersions, this axis of symmetry shifts.  Figure \ref{normsim2} displays the Normal simulation results for the case of $\mu_x=5, \mu_y=10, \theta_x=0.05$, and $\theta_y=0.025$.  Notice in this plot that sample sizes $n_x=80$ and $n_y=50$ cannot ensure a coverage probability of even 0.75 although the $X$ sample draws from the more moderate dispersion of $\theta=0.05$.\\

\noindent The two--sample Normal approximation appears to be considerably more robust than its one--sample counterparts.  Consider a one--sample case of $\mu=5, \theta=0.025$, and $n=100$ versus the two--sample case of $\mu_x=\mu_y=5$ and $\theta_x=\theta_y=0.025$ with $n_x=n_y=50$.  In either case, 100 i.i.d. data points are collected from the same experiment.  \citet{Shilane-negative-binomial-2010} showed that applying a Normal approximation to the one--sample case to estimate $\mu$ resulted in a coverage of 0.7802.  Meanwhile, the two--sample Normal confidence interval covered the mean $\mu_x-\mu_y$ with probability 0.9822.  (In this case, the one--sample data's mean is approximately Chi Square because $\mu=2n\theta$.  The Chi Square method covers $\mu$ at a rate of 0.9414.)  Even though the Normal approximation does not provide a good estimate to the one--sample data, its performance improves considerably in the two--sample case.  This trend is generally true across the entirety of the two--sample simulation experiments.  It appears that the two--sample Normal approximation is considerably more robust than its corresponding one--sample method.\\

\noindent When the Normal approximation fails to provide a strong coverage, the Bernstein method may be used as an alternative.  Figure \ref{bernsteinsim2} shows a broad range of values at which Bernstein confidence intervals improve upon the performance of the Normal approximation displayed in Figure \ref{normsim2}.  Furthermore, the Mixture Method that averages the Normal and Bernstein intervals shows that averaging the two methods results in confidence intervals that extend the range of values before the Bernstein method considerably over--covers the mean.\\

\noindent Table \ref{lengthtab} provides summary information for the median length of each method's confidence interval across all simulation experiments.  It also contains a summary of the difference in length between the Bernstein and Normal methods.  As expected, the Bernstein confidence intervals are considerably wider than the corresponding Normal intervals.  In many cases the Bernstein intervals are roughly double the length of the corresponding Normal interval.  When the Normal approximation performs poorly in terms of coverage, the wider Bernstein interval often provides an inference of higher quality.  When the Normal method performs well, the Bernstein confidence intervals will significantly over--cover the mean.  We can define the \emph{preference boundary} as the set of parameter values at which the Normal approximation overtakes the Bernstein method in terms of its coverage quality.  (For instance, this could be the point at which the Normal's coverage becomes closer to $1-\alpha$.)  In plots such as Figure \ref{normsim2}, with $\mu_x$, $\mu_y$, $\theta_x,$ and $\theta_y$ fixed, this boundary roughly takes the form of an ellipse defined on the sample sizes.  Within a neighborhood of this boundary, the Mixture method will outperform both the Bernstein and Normal methods.

\section{Discussion}\label{discussion}

\noindent At small values of $\theta$, Negative Binomial models produced highly skewed data that cause difficulties in drawing appropriate inferences about the mean.  The Normal approximation often performs poorly in one--sample settings.  However, Normal inferences on the two--sample mean difference $\mu_x-\mu_y$ are considerably more robust and can perform well even when neither individual sample is approximately Normal.  Tail probability bounds such as Bernstein's Inequality, along with the Normal--Bernstein Mixture method, provide complementary procedures.  Even under extreme dispersion at small sample sizes, the Bernstein method often performs well.  Although it is a conservative bound, Bernstein's Inequality emphasizes that the Normal approximation's confidence intervals are too narrow.  The Mixture method is intended to provide confidence intervals of intermediate length.  Indeed, an appropriately weighted combination of the Normal and Bernstein intervals can be constructed to produce a length anywhere in between the component results.\\

\noindent When at least one of the two samples follows a Gamma or Chi Square distribution, the Normal approximation was justified by a second--order Taylor series expansion of the cumulant function (the natural logarithm of the Laplace transform).  Future work could focus on further expanding this Taylor series.  We could examine the impact of third--order terms on the coverage of the Normal approximation and examine the drop--off in accuracy as $\alpha$ decreases.  Such an analysis would better justify inferences that run deeper into the tails of the distribution of $\bar X-\bar Y$ where the Normal approximation may become less accurate.\\

\noindent We further emphasize that the a--priori selection of the sample sizes $n_x$ and $n_y$ for controlled experiments is a difficult problem.  This is especially true in highly dispersed Negative Binomial models.  The simulation results suggest that equal dispersions imply that roughly equal sample sizes are preferable.  In other cases, some degree of imbalance would be preferred.  In selecting among the Normal, Bernstein, and Mixture methods, we offer the following limited guidelines:  The Bernstein method is typically preferred when the disparity in the sample sizes is large, especially for high dispersions.  In more moderate cases, the Normal approximation is the generally preferred method.  Finally, the Mixture method allows for the possibility of improvements over the Bernstein and Normal along the preference boundary.\\

\noindent The Bootstrap method was not included in the simulation study of Section \ref{sim} due to its burdensome computational requirements.  In the previous work of \citet{Shilane-negative-binomial-2010}, the Bootstrap BCA method was shown to produce similar results to the Normal approximation in one--sample settings.  Because the Normal approximation performs well in a greater variety of two--sample settings, the quality of two--sample Bootstrap inferences could be further investigated.

\bibliographystyle{chicago}
\bibliography{NB}

\end{document}